\documentclass{article}

\usepackage{amsmath}
\usepackage{amssymb}
\usepackage[mathcal,mathscr]{eucal}
\usepackage{eufrak}
\usepackage{epsfig}
\usepackage{amsthm}
\usepackage{graphicx}

\newcommand{\Ust}{U_{st}(t)}
\newcommand{\tr}{\textrm{tr}}
\newcommand{\vecol}[2]{\left(\begin{array}{c}
                      \hspace*{-1.5mm}#1\hspace*{-1.5mm} \\
                      \hspace*{-1.5mm}#2\hspace*{-1.5mm}
                       \end{array}\right)}
\newcommand{\vefil}[2]{\left(\begin{array}{cc}
                             \hspace*{-1.5mm}#1 & #2\hspace{-1.5mm} 
                             \end{array} \right)}
\newcommand{\matr}[4]{\left(\begin{array}{cc}
                              \hspace*{-1.5mm} #1 & #2\hspace*{-1.5mm} \\
                              \hspace*{-1.5mm} #3 & #4\hspace*{-1.5mm} 
                              \end{array} \right)}

\newcommand{\stint}[2]{\int_{0}^{t}#1(s)dW_{s}^{#2}}
\newcommand{\nn}{\nonumber}

\newtheorem{theor}{Theor.}[section]
\newtheorem{prop}{Prop.}[section]

\begin{document}


\title{\textbf{Lindblad Evolution and Stochasticity:}\\
\textbf{the Case of a Two-Level System}}
\author{D. Salgado\footnote{E-mail: david.salgado@uam.es}\hspace{0.5em} \& J.L. S\'{a}nchez-G\'{o}mez\footnote{E-mail:jl.sanchezgomez@uam.es}\\
Dpto. F\'{\i}sica Te\'{o}rica, Universidad Aut\'{o}noma de Madrid, Spain}
\maketitle
\begin{abstract}
We consider the simple hypothesis of letting quantum systems have an inherent random nature. Using well-known stochastic methods we thus derive a stochastic evolution operator which let us define a stochastic density operator whose expectation value under certain conditions satisfies a Lindblad equation. As natural consequences of the former assumption decoherence and spontaneous emission processes are obtained under the same conceptual scheme. A temptative solution for the preferred basis problem is suggested. All this is illustrated with a comprehensive study of a two-level quantum system evolution.
\end{abstract}

\vspace{1cm}


\section{Introduction}

Stochastic methods are lately  being profusely used in Quantum Mechanics in general and in Quantum Optics in particular to better study and analyze the physical processes taking place during the interaction between the system and the measuring apparatus. The conceptual framework is very clear, consisting in an open quantum system which is being continuously monitorized by an appropiate device. This situation is very well described using stochastic unravellings for the Lindblad equation of the system density operator, the most general expression of which for the diffusive case has been recently given in \cite{Diosi}. Needless to say, the presence of an observer is compulsory for orthodox quantum principles to be applied.

\bigskip

On the other hand, in the early years of Quantum Mechanics, the role of the observer was crucial to both justify the indeterministic nature of the results of a measurement upon quantum systems \cite{Dirac} and to understand the lack of quantum interference in a, say, double-slit experiment. Nowadays it is understood that this lack of interference is not strictly due to the presence of the observer, but to the possibility of knowing which path the particle followed  \cite{Englert}. Keeping orthodox principles \cite{Dirac} as far as possible, the role of the observer is then left as the only justification to set the indeterminism in the measurement process of quantum systems. 

\bigskip

In this article, we take under consideration the possibility of reducing once more the role of the observer and study the consequences of letting quantum systems have an intrinsic stochastic nature. In the same fashion, this hypothesis has already given rise to different stochastic evolution models (cf. \cite{Ghirardi} and references therein to previous works), the main purpose of which was to reproduce faithfully all nonrelativistic quantum predictions with the sole exception of quantum superpositions among macroscopically distinguishable states. By concentrating upon a very simple quantum system, namely a two-level system, we will show how this stochastic assumption allows us to make predictions only restricted to relativistic extensions of Quantum Mechanics (QED) or only achievable by resorting to the theory of open quantum systems. We will restrict ourselves to the nonrelativistic domain, i.e. we will keep on using state vectors to describe physical systems, something which is impossible to do in an attempt to merge relativistic principles with Quantum Theory.


\section{Motivation and General Framework}
\label{MotGenFram}

The main physical hypothesis we address here is the possibility of endowing the evolution operator of a quantum system with stochastic nature. To be concrete we will deal with a two-level quantum system in particular. This assumption has already been done in another places (cf. e.g. \cite{Giulini} for a general overview) and can be alternatively motivated in two ways. On one hand any system, whether quantum or classical, is continuously subjected to the external influence of its environment, however small the said influence may be. This standpoint is assumed in the program of decoherence, in fact it is its conceptual starting point \cite{Giulini}. Such external influences are, needless to say, uncontrollable, so we may encode them using stochastic methods. This way of proceeding is similar to for instance the modelling of the evolution of a Lorentz particle \cite{vanKamp}. In van Kampen's terminology it corresponds to what he calls \emph{external noise} which is due essentially to the huge amount of uncontrollable external factors that affect the system evolution. 

\bigskip

Alternatively, we may adopt the complementary point of view, stating that the stochasticity arises just as \emph{internal} noise (cf. again \cite{vanKamp}). This second standpoint is subtler and, in our opinion, much less intuitive but hopefully more accurate. Let us use a canonical example (which will be extensively studied below), a two-level quantum system in electromagnetic vacuum. Within the non-relativistic quantum-mechanical standpoint the evolution of this system is completely determined by its hamiltonian $H_{0}$ and since the latter is time-independent, the system is stationary, i.e. the energy does not change. Within QED framework, the description varies to include the effect of \emph{vacuum fluctuations}. These fluctuations, quantum in nature, drive the system to its ground energy level, thus producing \emph{spontaneous} decay. This situation can also be understood letting the system be modelled by stochastic evolution, rooting the stochasticity in this quantum vacuum. Since these fluctuations do not occur in real ordinary space, but in Hilbert space we use stochastic evolution operators to describe them. The term \emph{internal} is in this situation a bit dubious, since the electromagnetic vacuum can hardly be thought as internal to the two-level system. We should understand then \emph{internal} as making reference to the essence itself of the system. As a matter of fact this vacuum is only detectable through its effects on a quantum system, so in certain sense the evolution of the system should also contain such effects. 

\bigskip

From a mathematical point of view the question of the origin of the stochasticity is secondary and eveything reduces to find the form of such a stochastic evolution operator. So we proceed by substituting  $U(t)\rightsquigarrow U_{st}(t)$ and then investigating the form of $\Ust$. To do this we resort to a general decomposition theorem of real random variables \cite{Nualart} the generalization to (bounded) operator-valued random variables of which we take for granted.
 
\begin{theor}\label{DecompTh}
Let $X$ be an operator-valued random variable acting upon a N-dimensional Hilbert space $\mathfrak{H}$. Then there exist $N\times N$ operator-valued processes $v_{k}(t)$ such that 
\begin{equation}
X=\mathbb{E}X+\sum_{k=1}^{N^{2}}\int_{T}v_{k}(s)dW_{t}^{k},
\end{equation} 
where $\mathbb{E}$ denotes the expectation value with respect to the probability measure and $W_{t}^{k}$ are $n$ complex Wiener processes \footnote{Cf. e.g. \cite{Gardiner} for a general reference on stochastic methods.}.
\end{theor}
  
Expressing the latter integral as a function of the upper interval limit ($T\rightsquigarrow\mathbb{R}^{+}$) we may write as the more general form for a stochastic evolution operator 

\begin{equation}\label{StEvolOper}
U_{st}(t)=\mathbb{E}U_{st}(t)+\sum_{k=1}^{N^{2}}\int_{0}^{t}v_{k}(s)dW_{s}^{k},
\end{equation}

where in general the Wiener processes will have the covariance matrix given by

\begin{equation}
\mathbb{E}[W_{t}^{k}W_{t}^{k'*}]=a_{kk'}t.
\end{equation}

The connection with ordinary Quantum Mechanics is made by stating

\begin{equation}\label{DensOp}
\rho_{QM}(t)=\mathbb{E}\rho_{st}(t),
\end{equation}

where $\rho_{st}(t)$ is the system density operator induced by the previous stochastic evolution operator, namely

\begin{equation}
\rho_{st}(t)=\Ust\rho(0)U^{\dagger}_{st}(t).
\end{equation} 

Henceforward, we adopt the same assumptions as in the orthodox formalism, in particular we force trace-preserving evolution. This immediately imposes conditions on the absolute value of $\mathbb{E}\Ust$, as the following proposition shows:

\begin{prop}\label{ExpVal}
If $\rho_{QM}(t)$ is defined as (\ref{DensOp}) and $\tr\rho_{QM}(t)=\tr\rho_{QM}(0)$, then 
\begin{equation}\label{ExpValOpEvol}
\mathbb{E}\Ust=\exp(-iHt)\left(I-\sum_{nm=1}^{N^{2}}a_{nm}\int_{0}^{t}v_{m}^{\dagger}(s)v_{n}(s)ds\right)^{1/2}\hspace{-2mm},
\end{equation}
where $H$ is a selfadjoint operator.
\end{prop}

\begin{proof}
Cf. Appendix A.
\end{proof}

A few comments should be included. Firstly the selfadjoint operator $H$ is to be identified with the hamiltonian of the system. Secondly note, as expected, that if $v(t)\equiv 0$, then we recover the usual quantum-mechanical formalism, thus reinforcing the idea that new effects are contained only in the stochastic part of the evolution operator, leaving the hamiltonian partial evolution unchanged. Thirdly the square root imposes restriction upon $\sum_{nm}a_{nm}v_{m}^{\dagger}(t)v_{n}(t)$ since 

\begin{equation}
I-\sum_{nm=1}^{N^{2}}a_{nm}\int_{0}^{t}v_{m}^{\dagger}(s)v_{n}(s)ds\geq 0
\end{equation}

must hold for all $t\geq 0$. Finally $\Ust$ is not an unitary operator, unitarity\footnote{We are aware that the following condition only refers to partial isometry and not unitarity itself. We, for the time being, do not care about mathematical refinements.} is only obtained in stochastic average, i.e. 

\begin{equation}\label{StUnit}
\mathbb{E}[U^{\dagger}_{st}(t)\Ust]=I.
\end{equation}

This last feature is the crucial difference with other stochastic evolution schemes showing the same philosophy (specially those assumed in \cite{Partha} and \cite{Adler}). There, as in the scheme we propose here, it is assumed a basic stochastic evolution described by two operators, namely the Hamiltonian $H$ (for the deterministic part) and an \emph{ad hoc} operator $L$ which gives rise to the Lindblad operator \cite{Lindblad} after the whole calculation is carried on. Nevertheless the stochastic nature appears only as a modification of the evolution operator generator. For instance example 30.1 of \cite{Partha} proposes the \emph{ansatz} 

\begin{equation}
T(t)[\rho(0)]=\mathbb{E}[e^{iW_{t}L}\rho(0)e^{-iW_{t}L}],
\end{equation}

thus obtaining a Lindblad-type generator\footnote{The lack of a hamiltonian part can be viewed as a sort of hamiltonian-interaction picture election.}

\begin{equation}
\theta[X]=\frac{1}{2}\left\{ [LX,L]+[L,XL] \right\}.
\end{equation}

This is straightforwardly generalized to more than one Lindblad operator $L_{k}$. In a simpler and more intuitive form much the same it is arrived at in \cite{Adler}. They only confer stochastic nature to the evolution operator by adding random parts to its generator, thus obtaining some further restrictions on the final evolution, such as only selfadjoint Lindblad operators and consequently not achieving the desirable most general situation (any Lindblad operator). Parthasarathy has however provided a scheme to obtain general Lindblad evolution, namely he has proposed the evolution

\begin{equation}
T_{t}[X]=\mathbb{E}[U^{\dagger N(t)}XU^{N(t)}],
\end{equation}
 where $N(t)$ is Poisson process with intensity $\lambda$ and $L=\sqrt{\lambda}U$. Needless to say, though he arrives at a Lindblad generator $\frac{1}{2}\left\{ [L\cdot,L^{\dagger}]+[L,\cdot L^{\dagger}] \right\}$, the physical interpretation of such a stochastic evolution operator is rather evasive. So we find that to keep the intuitive approach of Adler, we must sacrifice some generality whereas on the other hand to arrive at such generality as Parthasarathy we lose physical intuition. In order to conjugate both characteristics, we have assumed a less restrictive position. 

\bigskip

However this generality does not drive us directly to a Lindblad evolution as we shall see. Some restrictions should be made on the added random parts $v_{k}(t)$. These restrictions possess, as the whole scheme, a rather intuitive physical interpretation. To see how the Lindblad evolution appears we focus on the differential version of (\ref{DensOp}). The starting point is the following

\begin{prop}
If $\rho(t)$ is defined as (\ref{DensOp}), then it satisfies the differential equation
\begin{equation}\label{DifEvolDensOp}
\dot{\rho}(t)=L_{t}[\rho(t)]+\widetilde{L}_{t}[\rho(0)],
\end{equation}

where for any $X$

\begin{subequations}
\begin{eqnarray}
L_{t}[X]&=&-i[H,X]+\nn\\
&+&\frac{1}{2}\sum_{nm=1}^{N^{2}}a_{nm}\left([\ell_{n}(t)X,\ell_{m}^{\dagger}(t)]+[\ell_{n}(t),X\ell_{m}^{\dagger}(t)]\right)\nn\\
& & \\
\widetilde{L}_{t}[X]&=&-\sum_{nm=1}^{N^{2}}a_{nm}L_{t}[\int_{0}^{t}v_{n}(s)Xv_{m}^{\dagger}(s)ds]
\end{eqnarray}
\end{subequations}
and by construction we have defined 

\begin{equation}\label{Lindblads}
\ell_{n}(t)=v_{n}(t)\mathbb{E}[U_{st}]^{-1}(t).
\end{equation}
\end{prop}

\begin{proof}
This result is straighforwardly obtained by deriving with respect to time the expression of $\rho_{QM}(t)$ (\ref{DensOp}) in which we use expression (\ref{ExpValOpEvol}) for $\Ust$ and identifying the previous definitions.
\end{proof}

\bigskip

The equation (\ref{DifEvolDensOp}) is almost a Lindblad differential equation. There are only two differences:

\begin{enumerate}
\item There is an extra term $\widetilde{L}_{t}[\rho(0)]$ which spoils the Markovianity. This on the other hand is also obtained in the more general quantum evolution of a subsystem in the orthodox formalism (cf. \cite{GFVKS} or \cite{Alicki}). The physical conditions to be met to guarantee Markovianity must be found.
\item The would-be Lindblad operators $\ell_{n}(t)$ are time-dependent. Though time-dependent Lindblad operators can be found in the literature (cf. \cite{Alicki} and references therein), this is not the usual case upon which we will focus. 
\end{enumerate}

Thus we must find the conditions under which on one hand $\widetilde{L}_{t}[\rho(0)]=0$ holds and on the other hand $\ell_{k}(t)\rightsquigarrow\ell_{k}$. In the next section we will comprehensively analyzed a two-level system to gain more physical insight into the question instead of trying to obtain the general conditions.

\bigskip

However notice that a first partial general result can be obtained as to the energy preserving or not preserving nature of the fluctuation part $v(t)$ once the second term $\widetilde{L}_{t}[\rho(0)]$ is shown to be negligible:

\begin{prop}
Both $[v(t),v^{\dagger}(t)]=0$ and $[H,v(t)]=0$ are independent sufficient conditions on their own to get energy conservation, i.e. 

\begin{equation}
\frac{dE(t)}{dt}\equiv\frac{d}{dt}\tr H\rho(t)=0.
\end{equation}
\end{prop}

\begin{proof}
Its elementary using (\ref{DifEvolDensOp}) with $\widetilde{L}_{t}[\rho(0)]=0$ and the cyclic property of the trace.
\end{proof}

This result may be useful to check whether we have a decaying or pumping interaction or on the contrary we just have a decohering but energy-preserving evolution (see below). 

\bigskip

Before working out the announced two-level quantum system example, we will briefly state where the main differences between previous stochastic quantum evolution models and our scheme are. Let us consider the Quantum State Diffusion model (cf. \cite{Percival}). This model corresponds to stochastic unravellings of the Lindblad equation under the hypothesis of state-vector normalization, which is implemented in the following way. If the stochastic state-vector satisfies the QSD equation

\begin{equation}
d|\psi_{st}\rangle=|v\rangle dt+|f\rangle d\xi,
\end{equation}

then the normalization preservation is achieved by forcing

\begin{equation}\label{OrtCond}
\langle\psi_{st}|f\rangle=0.
\end{equation}

As $|\psi_{st}\rangle$ must stochastically unravel the Lindblad equation, then

\begin{equation}\label{StocUnrav}
\rho(t)=\mathbb{E}[|\psi_{st}(t)\rangle\langle\psi_{st}(t)|].
\end{equation}

The QSD unravelling for the Lindblad equation with one Lindblad operator is

\begin{equation}
d|\psi_{st}\rangle=-iH|\psi_{st}\rangle dt+\left(\langle L^{\dagger}\rangle L-\frac{1}{2}L^{\dagger} L-\frac{1}{2}\langle L^{\dagger}\rangle\langle L \rangle\right)|\psi_{st}\rangle dt+(L-\langle L\rangle)|\psi_{st}\rangle d\xi,
\end{equation}
where $L$ denote the Lindblad operator entering the corresponding Lindblad equation and $\xi$ denotes white noise (cf. \cite{Percival}). Note that this equation is \textbf{non-local} and \textbf{non-linear}. The difference with the scheme proposed here is double. Firstly we do not assume the Lindblad equation as a starting hypothesis, we arrive at it only after finding some physical conditions (cf. below). Secondly we do not impose normalization of $|\psi_{st}\rangle$, we only impose the condition 

\begin{equation}
\mathbb{E}[|\langle\psi_{st}(t)|\psi_{st}(t)\rangle|^{2}]=1,
\end{equation}

i.e. normalization in stochastic average. To better appreciate these differences, we will explicitly write down the evolution equation for $|\psi_{st}\rangle$ derived from the evolution operator (\ref{StEvolOper}):

\begin{eqnarray}
d|\psi_{st}\rangle&=&(-iH-\frac{1}{2}\ell^{\dagger}(t)\ell(t))\mathbb{E}|\psi_{st}\rangle dt\nn\\
&+&v(t)|\psi(0)\rangle dW_{t}.
\end{eqnarray}

Notice that this equation is \textbf{linear} and, under appropiate elections of $v(t)$, \textbf{local}. Notice also the explicit appearance of $|\psi(0)\rangle$, in consonance with the most general evolution equation for an open quantum system \cite{GFVKS}. But QSD is not the unique diffusive stochastic unravelling for the Lindblad equation (cf. \cite{Diosi}). As a matter of fact it is included as a particular case of the stochastic evolution equation given by Di\'{o}si and Wiseman. Again, in their approach the Lindblad equation is assumed from the very beginning, which settles the first difference with the formalism proposed here. Secondly, as a consequence of the assumed conceptual framework (a quantum system being continuously monitorized), they impose the condition of purity conservation for the stochastic evolution, which gives rise to the main analytical difference, since in our approach $\Ust^{\dagger}\Ust\neq I$, as we have indicated above. Unitarity is only satisfied in stochastic average (cf. eq. (\ref{StUnit})). 

\bigskip

Differences can also be found with the work of Ghirardi, Pearle and Rimini \cite{Ghirardi}. There the Lindblad equation is not assumed from the beginning, but it is arrived at after imposing some other conditions. Among them, the most important one is, in our opinion, the election of the quantity to be containing the physical information of the quantum system. Their starting stochastic evolution equation is \textbf{linear} and (if appropiate elections are made) \textbf{local}, just as in our approach:

\begin{equation}
d|\psi\rangle=C|\psi\rangle dt + \mathbf{A}|\psi\rangle\cdot d\mathbf{B},
\end{equation}

 but they claim that, since this evolution does not preserve normalization, the physical information cannot be contained in $|\psi\rangle$. So they define another stochastic vector $|\phi\rangle$ by

\begin{equation}
|\phi\rangle\equiv\frac{1}{||\psi||}|\psi\rangle
\end{equation}

with probability function given by the probability function corresponding to $|\psi\rangle$ times their squared norm $||\psi||^{2}$, i.e.

\begin{equation}
q(\phi)=||\psi||^{2}p(\psi),
\end{equation}

where $q(\phi)$ denotes the probability associated to $|\phi_{st}\rangle$ and $p(\psi)$ denotes the probability associated to $|\psi_{st}\rangle$. This assumption leads immediately to the \textbf{nonlinear} stochastic differential equation 

\begin{equation}
d|\phi\rangle=\Big[-iH-\frac{\gamma}{2}(\mathbf{A}^{\dagger}-\mathbf{R}_{\phi})\cdot\mathbf{A}+\frac{\gamma}{2}(\mathbf{A}-\mathbf{R}_{\phi})\cdot\mathbf{R}_{\phi}\Big]|\phi\rangle dt + (\mathbf{A}-\mathbf{R}_{\phi})|\phi\rangle\cdot d\mathbf{B},
\end{equation}

where $\gamma$ denotes the non-null elements of the diagonal covariance matrix of the set of Wiener processes $\mathbf{B}$ and $\mathbf{R}_{\phi}\equiv\frac{1}{2}\langle\phi|\mathbf{A}+\mathbf{A}^{\dagger}|\phi\rangle$. Now the difference is rooted in the assumption that in our approach nonnormalization does not suppose a major nuisance, since we claim that the physical information of the system is contained in 

\begin{equation}\label{DensOp2}
\rho(t)=\mathbb{E}[|\psi_{st}\rangle\langle\psi_{st}|],
\end{equation}

so that the probability of finding a system in the state $P_{\sigma}=|\sigma\rangle\langle\sigma|$ will be

\begin{equation}
\tr\rho P_{\sigma}=\mathbb{E}[|\langle\psi_{st}|\sigma\rangle|^{2}].
\end{equation}

And it is the physical quantity (\ref{DensOp2}) which is to be normalized. Thus we do not need to resort to some other normalized process $|\phi\rangle$.

\bigskip

Finally, though being very close in spirit, some differences can also be remarked with the work of Gisin \cite{Gisin}. Here the objective is to classify all the pure state valued stochastic differential equations in $\mathbb{C}^{2}$ such that the corresponding density matrix follows a quantum dynamical semigroup evolution. The first formal difference stems form the fact that he describes the stochastic evolution in terms of the ``height'' and azimutal angle of the corresponding point in the Bloch Sphere, thus preventing its analysis from being generalized to higher-level systems in a straightforward fashion. Moreover he only focus on pure states in contrast to our density matrix formalism, which also embraces mixture states. Last of all, again as in previous commented models, the evolution equation for the density operator is assumed from the beginning, though the restriction of complete positivity is not guaranteed in his analysis.

\section{Non-perturbed Two-level Systems}
\label{NonPert}

We begin by considering non-perturbed systems, i.e. with hamiltonian of the form 

\begin{equation}
H_{0}=\matr{E_{+}}{0}{0}{E_{-}}
\end{equation} 

The reason to focus only on non-perturbed hamiltonians will be clear in the next section. As usual (cf. e.g. \cite{Peng}) we may choose $E_{+}+E_{-}=0$, thus we write

\begin{equation}
H_{0}=\hbar\omega_{0}\matr{1}{0}{0}{-1}
\end{equation} 

As a canonical example we may consider an atom in electromagnetic vacuum. The random part of the stochastic evolution operator may be written in the energy eigenvector basis as 

\begin{eqnarray}
\matr{\stint{\lambda_{11}}{1}}{0}{0}{0}&+&\matr{0}{\stint{\lambda_{12}}{2}}{0}{0}\nn\\\label{CompWis}+\matr{0}{0}{\stint{\lambda_{21}}{3}}{0}&+&\matr{0}{0}{0}{\stint{\lambda_{22}}{4}}
\end{eqnarray}

This is the most general form. To clearly understand the physical meaning of each of the previous four terms we will start by considering them one by one. 

\subsection{Single Couplings: Decohering, Decaying and Pumping Factors}

Thus let us study the evolution induced by a stochastic evolution operator of the form

\begin{eqnarray}
\Ust&=&e^{-iH_{0}t}\left[I-\int_{0}^{t}v^{\dagger}(s)v(s)ds\right]^{1/2}\nn\\
&+&\int_{0}^{t}v(s)dW_{s},
\end{eqnarray}

where the stochastic expectation value has already been written to preserve the trace, $v(t)$ is one of the previous $\lambda_{ij}(t)E_{ij}$'s ($\{E_{ij}\}$ being the canonical basis of $\mathcal{M}_{2}(\mathbb{C})$) and $W_{t}$ is the complex standard Wiener process. As a first result we elementarily arrive at $\widetilde{L}_{t}[\rho(0)]=0$ for any election of $v(t)$, so Markovianity is straightforwardly achieved. To obtain time-independent Lindblad operators, we insert sucessively each $v(t)$ in the definition of $\ell(t)$ (eq. (\ref{Lindblads})). Forcing time-independency sets an ordinary differential equation for $|\lambda_{ij}(t)|^{2}$ whose solution leaves as the only option for $\lambda_{ij}(t)$ the expression

\begin{equation}
\lambda_{ij}(t)=\gamma^{1/2}\exp\left(-\frac{\gamma t}{2}\mp i\omega_{0}t\right).
\end{equation}

The Lindblad operators produced in this way are of the form 

\begin{equation}
\ell=\gamma E_{ij}
\end{equation}

The physical interpretation is rather intuitive. If for instance we choose $v(t)=\lambda_{11}(t)E_{11}$, then we obtain an energy-preserving decohering evolution\footnote{As usual $\rho_{ij}(t)$ denote the $ij$th entry of the density matrix $\rho(t)$.}
                                              , i.e. 

\begin{equation}\label{OnlyDecoh}
\frac{d\rho(t)}{dt}=-\frac{i}{\hbar}[H_{0},\rho(t)]+\frac{\gamma}{2}\matr{0}{-\rho_{12}(t)}{-\rho_{21}(t)}{0}.
\end{equation}

\begin{figure}[htb]
\begin{center}
\includegraphics[width=8.5cm]{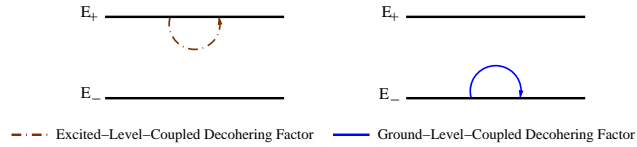}
\caption{Representation of energy-preserving decohering factors in the stochastic evolution of two-level quantum system.\label{DecohFactors}}
\end{center}
\end{figure}

This means that $W^{1}(t)$ represents a vacuum-system interaction which does not change the energy of the system, but that introduces decohering factors. Notice that exactly the same equation (\ref{OnlyDecoh})  would have been obtained had we chosen $v(t)=\lambda_{22}(t)E_{22}$. The difference stems from the fact that whereas $W^{1}(t)$ represents an interaction through coupling to the excited energy level, $W^{4}(t)$ represents a similar interaction but coupled to the ground energy level. Graphically the situation can be represented as in Fig. \ref{DecohFactors}.

\bigskip

This same interpretation can be carried over to the other two terms. For instance, for $v(t)=\lambda_{21}(t)E_{21}$ the final evolution equation is

\begin{equation}\label{SponEmiss}
\frac{d\rho(t)}{dt}=-\frac{i}{\hbar}[H_{0},\rho(t)]+\frac{\gamma}{2}\matr{-2\rho_{22}(t)}{-\rho_{12}(t)}{-\rho_{21}(t)}{2\rho_{22}(t)}.
\end{equation}
 
\begin{figure}[htb]
\begin{center}
\includegraphics[width=6cm, height=5cm]{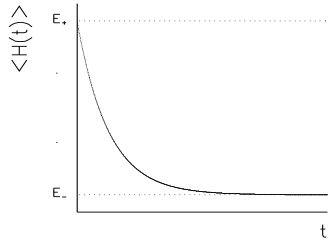}
\vspace{-1.0cm}
\caption{Spontaneous Decay of a Two-level Atom.\label{SponDecay}}
\end{center}
\end{figure}

\begin{figure}[htb]
\begin{center}
\includegraphics[width=8.5cm]{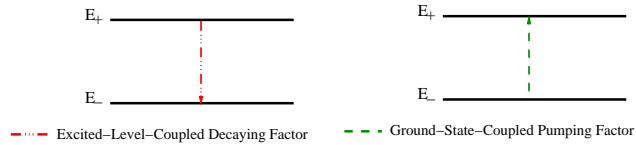}
\caption{Representation of energy-nonpreserving decohering factors in the stochastic evolution of two-level quantum system.\label{DecPumpFactors}}
\end{center}
\end{figure}

Now besides the off-diagonal decohering terms a new effect appears: the system \emph{spontaneously} decays into the ground state. This behaviour is indeed exactly what is obtained by orthodox means (cf. e.g. \cite{Peng} and Fig. \ref{SponDecay}). The reverse behaviour, as expected, is achieved if we choose $v(t)=\lambda_{12}(t)E_{12}$. In these two cases then the vacuum-system interaction is understood as an energy-non-preserving (decaying or pumping) interaction, depending upon whether the coupling takes place through the excited or the ground state. Pictorially we may represent these effects as in Fig. \ref{DecPumpFactors}. The whole results are collected in table \ref{OneStPart} for each of the previous stochastic parts.

\bigskip

\begin{table*}
\begin{center}
\begin{tabular}{c||c||c}
\textsf{Stochastic Term} & $\lambda_{ij}(t)$ & \textsf{Lindblad Operator} \\ \hline\hline
 & & \\
$\lambda_{11}(t)\matr{1}{0}{0}{0}$ & $\lambda_{11}(t)=\gamma^{1/2}\exp\left(-\frac{\gamma t}{2}-i\omega_{0}t\right)$ & $\gamma\matr{1}{0}{0}{0}$ \\
 & & \\ 
 & & \\ \hline
 & & \\
$\lambda_{12}(t)\matr{0}{1}{0}{0}$ & $\lambda_{12}(t)=\gamma^{1/2}\exp\left(-\frac{\gamma t}{2}+i\omega_{0}t\right)$ & $\gamma\matr{0}{1}{0}{0}$ \\
 & & \\
 & & \\ \hline
 & & \\
$\lambda_{21}(t)\matr{0}{0}{1}{0}$ & $\lambda_{21}(t)=\gamma^{1/2}\exp\left(-\frac{\gamma t}{2}-i\omega_{0}t\right)$ & $\gamma\matr{0}{0}{1}{0}$ \\
 & & \\
 & & \\ \hline
 & & \\
$\lambda_{22}(t)\matr{0}{0}{0}{1}$ & $\lambda_{22}(t)=\gamma^{1/2}\exp\left(-\frac{\gamma t}{2}+i\omega_{0}t\right)$ & $\gamma\matr{0}{0}{0}{1}$ \\
 & & \\\hline\hline
\end{tabular} 
\end{center}
\caption{Lindblad Operators and Stochastic Parts for single vacuum-system couplings.\label{OneStPart}} 
\end{table*}


Notice that Markovianity is obtained with no further restrictions upon the form of the stochastic parts $v_{k}(t)$'s. The final form of their entries is calculated from the time-independence condition for the Lindblad operators. On the contrary, some physical conditions are to be met when two Wiener processes are present.

\subsection{Double couplings}

The next natural step is to include one more Wiener process in the stochastic evolution, so we must investigate the options

\begin{eqnarray}
\Ust\hspace{-2mm}&=&\hspace{-2mm}e^{-iH_{0}t}\hspace{-1mm}\left[I-\sum_{ij=1}^{2}a_{ij}\int_{0}^{t}v_{j}^{\dagger}(s)v_{i}(s)ds\right]^{1/2}\hspace{-4mm}+\nn\\
&+&\int_{0}^{t}v_{1}(s)dW_{s}^{1}+\int_{0}^{t}v_{2}(s)dW_{s}^{2}\hspace{-3mm}
\end{eqnarray}

where $v_{j}(t)$ will alternately be each $\lambda_{ij}(t)E_{ij}$. The calculations proceed exactly in the same spirit as before, with the exception that now the covariance matrix $\mathbb{E}[W_{t}^{n}W_{t}^{m*}]=a_{nm}t$ must be taken into account. The results are contained in table \ref{TwoStPart}.

\bigskip

\begin{table*}
\begin{tabular}{c||c||c}
\textsf{Stochastic Terms} & $a_{ij};\lambda_{ij}(t)$ & \textsf{Lindblad Operators} \\ \hline\hline
 & $a_{12}=0$ & \\
$\matr{\lambda_{11}(t)}{0}{0}{0}\qquad \matr{0}{\lambda_{12}(t)}{0}{0}$ &$\lambda_{11}(t)=\sqrt{\gamma_{1}}\exp\left(-\frac{a_{11}\gamma_{1}}{2}t-i\omega_{0}t\right)$  & $\gamma_{1}\matr{1}{0}{0}{0}\qquad \gamma_{2}\matr{0}{1}{0}{0}$ \\
 & $\lambda_{12}(t)=\sqrt{\gamma_{2}}\exp\left(-\frac{a_{22}\gamma_{2}}{2}t+i\omega_{0}t\right)$ & \\ 
 & & \\ \hline
 & & \\
 & & It's impossible to have both a \\
$\matr{\lambda_{11}(t)}{0}{0}{0}\qquad \matr{0}{0}{\lambda_{21}(t)}{0}$ &  & preserving and a nonpreserving \\
 & &  factor coupled to the same level. \\ 
 & & \\ \hline
 & & \\
 & $a_{14}=0$ & \\
$\matr{\lambda_{11}(t)}{0}{0}{0}\qquad \matr{0}{0}{0}{\lambda_{22}(t)}$ & $\lambda_{11}(t)=\sqrt{\gamma_{1}}\exp\left(-\frac{a_{11}\gamma_{1} t}{2}-i\omega_{0}t\right)$ & $\gamma_{1}\matr{1}{0}{0}{0}\qquad \gamma_{4}\matr{0}{0}{0}{1}$ \\
 &$\lambda_{22}(t)=\sqrt{\gamma_{4}}\exp\left(-\frac{a_{44}\gamma_{4} t}{2}+i\omega_{0}t\right)$ & \\ 
 & & \\ \hline
 & & \\
& & It's impossible to obtain a \\
$\matr{0}{\lambda_{12}(t)}{0}{0}\qquad \matr{0}{0}{\lambda_{21}(t)}{0}$ &  & Markovian evolution \\
 & &  for any initial $\rho(0)$. \\ 
 & & \\ \hline
 & & \\
 & & It's impossible to have both a \\
$\matr{0}{\lambda_{12}(t)}{0}{0}\qquad \matr{0}{0}{0}{\lambda_{22}(t)}$ &  & preserving and a nonpreserving \\
 & &  factor coupled to the same level. \\ 
 & & \\ \hline
 & & \\
 & $a_{34}=0$ & \\
$\matr{0}{0}{\lambda_{21}(t)}{0}\qquad \matr{0}{0}{0}{\lambda_{22}(t)}$ & $\lambda_{21}(t)=\gamma_{3}^{1/2}\exp\left(-\frac{a_{33}\gamma_{3} t}{2}-i\omega_{0}t\right)$ & $\gamma_{3}\matr{1}{0}{0}{0}\qquad \gamma_{4}\matr{0}{0}{0}{1}$ \\
 &$\lambda_{22}(t)=\gamma_{4}^{1/2}\exp\left(-\frac{a_{44}\gamma_{4} t}{2}+i\omega_{0}t\right)$ & \\ 
 & & \\ \hline
\end{tabular}
\caption{Lindblad Operators and Stochastic Parts for double vacuum-system couplings.\label{TwoStPart}} 
\end{table*}

Some comments are to be made. First the impossibilty of having a double coupling to the same level is clearly shown (rows 2 and 5), which is quite natural from a physical point of view. Perhaps the most appealing feature is the necessary uncorrelation between the two Wiener processes expressed through the condition $a_{ij}=0$. There's an immediate interpretation of this condition in the basis of Markovian evolution. Physically  Markovianity  is characterized by the fact that the evolution of a system from one initial time to a final time is exactly the same as the evolution from that initial time to an intermediate time and then from the latter to the final instant, \emph{whatever} that intermediate time is. This strongly suggests the idea that the evolution process is made up of very tiny and identical contributions of infinitesimal duration, so that the final evolution is just the whole contribution of each infinitesimal part. The common view states that such a system lacks of memory, i.e. the state to which it evolves only depends on its present state and never in the previous ones. These two standpoints are complementary. In our case if the Wiener processes partially driving the system evolution were correlated, then this evolution partition into tiny equal parts would be impossible, thus neglecting the possibility of Markovianity. The final form of the matrix entries $\lambda_{ij}(t)$'s are obtained requiring time-independence as before.

\subsection{Triple and Quadruple Couplings}

The considerations taken up before lead us immediately to neglect any possibility of establishing  triple or quadruple couplings, since each energy level can only interact with the electromagnetic vacuum through one single Wiener process. Only in case of more complex systems we may combine in a variety of different ways triple, quadruple and higher-order couplings.

\section{Perturbed Systems}

The generalization of the previous formalism to perturbed, i.e. interacting systems poses relevant questions as to the adopted conceptual scheme. Naively we may try to investigate the evolution produced by an operator of the form 

\begin{eqnarray}
\Ust\hspace{-2mm}&=&\hspace{-2mm}\mathcal{T}e^{-i\int_{0}^{t}H(s)ds}\left[I-\int_{0}^{t}v^{\dagger}(s)v(s)ds\right]^{1/2}\hspace{-3mm}+\nn\\
\label{UstNonDiag}&+&\int_{0}^{t}v(s)dW_{s},\hspace{-1cm}
\end{eqnarray}

where $H(t)$ may be nondiagonal, as e.g. that expressing the interaction with an electromagnetic wave\footnote{The rotating wave approximation has already been assumed.} 

\begin{equation}
H(t)=\matr{\hbar\omega_{0}}{\mathcal{E}e^{-i\omega t}}{\mathcal{E^{*}}e^{i\omega t}}{-\hbar\omega_{0}}.
\end{equation}

Hopefully this as before should drives us to a differential equation for the density operator with the form

\begin{equation}
\frac{d\rho(t)}{dt}=-i[H(t),\rho(t)]+\frac{1}{2}\left\{[\ell\rho(t),\ell^{\dagger}]+[\ell,\rho(t)\ell^{\dagger}]\right\}.
\end{equation}

In particular one should be able to reproduce Rabi oscillations with decay (cf. e.g. \cite{Loudon}), but this is impossible as the following proposition states:

\begin{prop}\label{Prop4}
  Let $X\in\mathcal{M}_{2}(\mathbb{C})$ be of the form $X=v\rho v^{\dagger}$ where $\rho\in\mathcal{M}_{2}(\mathbb{C})$ is a density matrix and $v\in\mathcal{M}_{2}(\mathbb{C})$ is arbitrary. If in an arbitrary common basis, $H$ and $\ell$ are expressed as $$H=\matr{h_{11}}{h_{12}}{h_{21}}{h_{22}}\qquad \ell=\matr{\ell_{11}}{\ell_{12}}{\ell_{21}}{\ell_{22}},$$then one of the following necessary conditions to have $$L[X]\equiv-i[H,X]+\frac{1}{2}\left\{[\ell X,\ell^{\dagger}]+[\ell,X\ell^{\dagger}]\right\}=0$$ $H \textrm{ selfadjoint},\ell\textrm{ arbitrary}$ for all $\rho$ must hold
\begin{subequations}
\begin{enumerate}
\item If $v=\matr{\beta_{1}}{\beta_{2}}{0}{0}$, then \begin{equation}\label{case1}\ell_{12}\ell_{11}^{*}=i2h_{12}\qquad\ell_{21}=0\end{equation}
\item If $v=\matr{0}{0}{\beta_{1}}{\beta_{2}}$, then \begin{equation}\label{case2}\ell_{21}\ell_{22}^{*}=i2h_{21}\qquad\ell_{12}=0\end{equation}
\item If $v=\matr{\beta_{1}}{\beta_{2}}{\beta_{3}}{\beta_{4}}\qquad\beta_{k}$'s not corresponding to the previous cases, then \begin{eqnarray}
                   &  \alpha_{1}\Lambda_{11}+\alpha_{2}\Lambda_{21}+\alpha_{3}\Lambda_{31}+\alpha_{4}\Lambda_{41}=0 & \nn \\                             &  \alpha_{1}\Lambda_{12}+\alpha_{2}\Lambda_{22}+\alpha_{3}\Lambda_{32}+\alpha_{4}\Lambda_{42}=0 & \nn \\
                   & & 
                     \end{eqnarray}
where the $\alpha_{i}$'s and the $\Lambda_{ij}$'s are complicated expressions concerning the elements of $H$ and $\ell$ (cf. Appendix \ref{AppendProp4}).
\end{enumerate}
\end{subequations}
\end{prop}

\begin{proof}
Cf. Appendix \ref{AppendProp4}.
\end{proof}

To get a decay process we only need case 2-- eq. (\ref{case2})-- since the Lindblad operator now reads $\ell=\gamma E_{21}$ which is only possible if $v$ is of type $\left(\begin{smallmatrix}0&0\\\beta_{1}&\beta_{2}\end{smallmatrix}\right)$ which following (\ref{case2}) makes the Lindblad components time-dependent and the Hamiltonian diagonal.

\bigskip

Far from invalidating the whole formalism this, in our opinion, suggests the correct interpretation to be assumed in the introduction of the stochastic nature of the evolution operator: the former scheme must only be applied to closed systems. The question then is \emph{why is it possible to apply this formalism only to traditionally closed systems?} The answer can be inspired in the canonical example used so far. In an orthodox description, the presence of just a single photon changes the whole picture, since now there's a \textbf{real} interaction between the electromagnetic field and the atom. In the former case, with no photon, i.e. with electromagnetic vacuum, the notion of interaction does not appear in a natural way, since there's nothing the atom interacts with. But this does not mean that an open (interacting) system may not show stochastic behaviour, what this means is that to obtain the correct stochastic evolution operator for an open system, we must apply the previous formalism to the system+environment compound and then trace out the latter's degrees of freedom. In the general case this is nearly impossible since it supposes to be able to control all environmental degrees of freedom. To illustrate this idea and check whether the qualitative behaviour described in the previous section is still valid, we will work out a simple and manageable example, namely a two-level atom described by a Jaynes-Cummings Hamiltonian

\begin{equation}
H_{JF}=\omega_{0}S_{z}+\omega a^{\dagger}a+\epsilon(a^{\dagger}S_{-}+aS_{+}).
\end{equation}

This represents an atom interacting only with one single mode of the electromagnetic field. We will consider only a multiple energy-preserving interaction, i.e. for the whole atom+field system we consider a stochastic evolution operator of the form\footnote{We denote $\lambda_{j}^{\pm}\circ W_{j}^{\pm}(t)\equiv\int_{0}^{t}\lambda_{j}^{\pm}(s)dW_{j}^{\pm}(s)$.} 

\begin{eqnarray}
\Ust&=&\mathbb{E}\Ust+\sum_{n=0}^{\infty}\lambda_{n+1}^{\pm}\circ W_{n+1}^{\pm}(t)|u_{n+1}^{\pm}\rangle\langle u_{n+1}^{\pm}|+\nn\\
&+&\lambda_{0}^{-}\circ W_{0}^{-}(t)|u_{0}^{-}\rangle\langle u_{0}^{-}|,
\end{eqnarray}

where $\lambda_{n}^{\pm}(t)\equiv\sqrt{\gamma_{n}^{\pm}}\exp\left[-\frac{\gamma_{n}^{\pm}}{2}t\mp i\omega_{n}^{\pm}t\right]$ and $\mathbb{E}\Ust$ has the corresponding form given by Prop. \ref{ExpVal}.

\bigskip

The previous formalism yields the following solution for the density matrix entries:

\begin{subequations}
\begin{eqnarray}
\label{OffDiag}\rho_{nm}^{\pm\pm}(t)&=&\exp\negthinspace\left[-i(\omega_{n}^{\pm}-\omega_{m}^{\pm})t-\frac{1}{2}(\gamma_{n}^{\pm}+\gamma_{m}^{\pm})t\right]\rho_{nm}^{\pm\pm}(0) \textrm{ if $n\neq m$ or different signs,}\\
\label{Diag}&=&\rho_{nn}^{\pm\pm}(0) \hspace{1cm}\textrm{otherwise,} 
\end{eqnarray}
\end{subequations}
which is the similar decohering behaviour previously shown by the single atom. But we are interested in the atom itself and not in the whole system, so we must trace out the field degrees of freedom, i.e.

\begin{equation}\label{ParTrac}
\rho_{A}(t)=\tr_{F}\rho_{AF}(t).
\end{equation}

To be concrete we will focus on two particular situations. First we will consider an initially correlated or uncorrelated excited atom and $n$ photons with energy $\hbar\omega$ ($n\neq 0$). Later we will study the case of no photon ($n=0$). Under these hypotheses the initial atom+field density matrix in the $|\pm,m\rangle$ basis for the first case is given by

\begin{equation}\label{InitCond}
\rho(\cdot,\times;pq)=\delta_{\cdot +}\delta_{\times +}\delta_{pn}\delta_{qn},
\end{equation} 

where by $\cdot,\times$ we denote any of the $\pm$ signs. The partial trace (\ref{ParTrac}) drives us through a tedious though elementary calculation to the result 

\begin{eqnarray}
\rho_{A}&=&\left[\sum_{k=0}^{\infty}\left(\rho_{k+1 k+1}^{++}\cos^{2}\theta_{k+1}-(\rho_{k+1 k+1}^{+-}-\rho_{k+1 k+1}^{-+})\cos\theta_{k+1}\sin\theta_{k+1}+\rho_{k+1 k+1}^{--}\sin^{2}\theta_{k+1}\right)\right]|+\rangle\langle+|+\nn\\
        &+&\Bigg[\sum_{k=1}^{\infty}\left(\rho_{k+1 k}^{++}\cos\theta_{k+1}\sin\theta_{k}+\rho_{k+1 k}^{+-}\cos\theta_{k+1}\cos\theta_{k}-\rho_{k+1 k}^{-+}\sin\theta_{k+1}\sin\theta_{k}-\right.\nn\\
&-&\left.\rho_{k+1 k+1}^{--}\sin\theta_{k+1}\cos\theta_{k}\right)+\rho_{10}^{+-}\cos\theta_{1}-\rho_{10}^{--}\sin\theta_{1}\Bigg]|+\rangle\langle-|+\nn\\
        &+&\Bigg[\sum_{k=1}^{\infty}\left(\rho_{k k+1}^{++}\cos\theta_{k+1}\sin\theta_{k}-\rho_{k k+1}^{+-}\sin\theta_{k+1}\sin\theta_{k}+\rho_{k k+1}^{-+}\cos\theta_{k+1}\cos\theta_{k}-\right.\nn\\
&-&\left.\rho_{k k+1}^{--}\sin\theta_{k+1}\cos\theta_{k}\right)+\rho_{01}^{-+}\cos\theta_{1}-\rho_{01}^{--}\sin\theta_{1}\Bigg]|-\rangle\langle+|+\nn\\
        &+&\Bigg[\sum_{k=1}^{\infty}\left(\rho_{k+1 k+1}^{--}\sin^{2}\theta_{k+1}+(\rho_{k+1 k+1}^{+-}+\rho_{k+1 k+1}^{-+})\cos\theta_{k+1}\sin\theta_{k+1}+\rho_{k+1 k+1}^{--}\cos^{2}\theta_{k+1}\right)+\nn\\
\label{PartDens}&+&\rho_{00}^{--}\Bigg]|-\rangle\langle-|,
\end{eqnarray}

where $\rho_{pq}^{\cdot\times}$ are the density matrix components in the $\{|u_{n+1}^{\pm}\rangle,|u_{0}^{-}\rangle\}$ basis. Notice that all these matrix density entries in (\ref{PartDens}) are time-dependent, the time dependence being given by (\ref{OffDiag}) and (\ref{Diag}). Indeed as a result of (\ref{OffDiag}) and (\ref{Diag}), and the initial conditions (\ref{InitCond}) the final atom state after a time $t\gg T\equiv \max\{(\gamma_{i}+\gamma_{j})^{-1}\}$ has elapsed is a $2\times 2$ matrix given in the usual $|\pm\rangle$ basis by

\begin{equation}
\rho_{A}(t\gg T)\simeq\left(\begin{smallmatrix}\cos^{4}\theta_{n+1}+\sin^{4}\theta_{n+1}&0\\0& 2\cos^{2}\theta_{n+1}\sin^{2}\theta_{n+1}\end{smallmatrix}\right),
\end{equation}

from where we can explicitly follow the decoherence suffered by the atom, and the probabilities of remaining and decaying in the excited and ground states respectively exactly coincide with the quantum ones. As initially chosen (in the election of the stochastic part), the energy of atom+field is conserved, so we only obtain decoherence in the compound system. However the energy of the system may vary, as its reduced density matrix shows ($\rho_{A}(--)\neq 0$), this energy variation being due to the interaction between the field and the atom. 

\bigskip

The second situation in which there's no photon present ($n=0$) is included to check consistency with the case of a single atom --see section \ref{NonPert}. Now a quantitative difference should be expected, since we are only taking into account one single mode. However the qualitative behaviour, i.e. the \emph{spontaneous decay} process should also be obtained. The same density matrix form for large $t$ is as a matter of fact obtained in a similar calculation

\begin{equation}
\rho_{A}(t\gg T)\simeq\left(\begin{smallmatrix}\cos^{4}\theta_{1}+\sin^{4}\theta_{1}&0\\0& 2\cos^{2}\theta_{1}\sin^{2}\theta_{1}\end{smallmatrix}\right).
\end{equation}
As expected we do not obtain the same quantitative behaviour as in (\ref{SponEmiss}). However a qualitative spontaneous emission is obtained, since starting from a system completely in the excited state (cf. eq. (\ref{InitCond})) we have found that there's nonnull probability of finding it in the ground level even though there's no photon present in the field.

\section{Vacuum Interactions vs. Stochastic Nature}

So far we have kept ourselves under the conceptual orthodoxy derived from the Quatnum Field Theory formalism in general, and the Quantum Electrodynamics in particular. But a natural question can be raised as to the conceptual framework which may be attached to the previous section scheme. Is it possible to found the motivation of use of stochastic methods on some other vacuum-independent notions? This is, we are aware, a delicate question, since it reaches the very interpretation of Quantum Mechanics itself. But nothing is further from our intention than providing a brand new interpretation of Quantum Mechanics. We only restrict ourselves to consider the possibility of merging together the stochastic nature of quantum systems (derived from the projection postulate) with the usual unitarity evolution. In this process it is straightforward to convince oneself that the original Dirac's argument (cf. \cite{Dirac}) is still valid, i.e. any linear superposition of states must be conserved during the evolution of the system, hence a linear operator can be defined which carries the initial states onto the evolved states at time $t$.

\bigskip

In QED it is the virtual quantum fluctuations which drive the system from initial excited states into its ground state. This process is equivalent to recognizing a patent stochasticity in the whole system (what other meaning can \emph{fluctuations} bear?), a system which is necessarily compound (atom+field), even though there's no photon present in the field. Is it not much easier and more natural to assume that quantum systems possess an inherent random nature independent of outer entities (fields, environments, etc)? Instead of rooting from a fundamental point of view the stochasticity of quantum systems in external incontrollable perturbations (observations, measurements), cannot we assert that this randomness is due to their quantum nature itself? Another way of stating this hypothesis is by realizing that in orthodox Quantum Mechanics, indeterminism is only present in the act of measuring, thus neglecting this indeterminism in the nature of the quantum systems itself (cf. \cite{Dirac}, page 108). As a matter of fact quantun systems which are not observed are deterministic. Nevertheless our experience in the laboratory clearly demands an indeterministic behaviour of these systems. Is it not more natural just to claim that these quantum systems are indeterministic themselves, with no resorting to external observations or interventions? Note that this does not suppose a major departure from quantum orthodoxy, so we do not lose any result characteristic of Quantum Mechanics. Indeed the stochasticity, present in the formalism through the operator $v(t)$ (cf. above), can be chosen to slightly modify the usual hamiltonian evolution. Suppose that the effects of the stochasticity are small compared to the hamiltonian evolution, i.e. we may write 

\begin{equation}
\left[I-\int_{0}^{t}v^{\dagger}(s)v(s)ds\right]^{1/2}\approx I-\frac{1}{2}\int_{0}^{t}v^{\dagger}(s)v(s)ds,
\end{equation}

where $v^{\dagger}(t)v(t)$ is of higher order than $H$ in the norm sense. Then we may write for the density operator 

\begin{equation}
\rho(t)=\rho_{QM}(t)+\widetilde{\rho}_{st}(t).
\end{equation}

The usual well-known quantum results are contained in the first part, whereas new effects are only present in the stochastic part. As we have seen for two-level quantum systems, these new effects amount to introducing either energy-preserving terms which only show decoherence (only achieved by resorting to other systems in orthodox Quantum Mechanics) or energy-nonpreserving terms which in a natural way show spontaneous decay phenomena also without resorting to external interventions.

\bigskip

On the other hand a mathematical scheme accounting for the stochastic nature of quantum systems is provided, as we have implicitly done, by letting theorem \ref{DecompTh} be applied to each evolution operator component. Several comments should be remarked. First an elaborate argument should be offered for the cited theorem to be applied. This argument reads as follows. The stochasticity of quantum systems, as it is expressed by the projection postulate, roots on the impossibility of determining the evolution of a quantum system at a given instant $t$ (typically when a measurement upon it is performed). At most we can know the probabilities of evolution from that state to the subsequent ones (typically the eigenvectors of the observable to be measured). The need for the presence of an observer has been the origin of neverending discussions since its very introduction. The mathematical way to deal with this kind of stochastic objects in general is to use random variables, which are nothing more than deterministic (usual) variables with a probability measure associated to them. So cannot quantum states, i.e. Hilbert-space vectors be also attached probability measures? That's all we have done. Once we have assumed such a hypothesis, the foregoing construction rests on mathematical results. If a linear operator tranforms quantum states into quantum states, then a stochastic linear operator will also transform stochastic quantum states into stochastic quantum states, hence endowing operators with an intrinsic random linear nature. This randomness can be expressed, as we have done, componentwise, so turning each operator (complex) entry into a complex random variable. Next step was then to apply theorem \ref{DecompTh} to these complex random variables.

\bigskip

But a subtlety must be considered. When the whole process is built componentwise (cf. eq.(\ref{CompWis})), one may (and should) inquiry \emph{which basis is to be used}? This surpringly drives us to the \emph{preferred basis problem}. But now instead of trying to provide more or less complicate theoretical arguments, we may allow Nature to freely choose the basis. To know what basis Nature chooses we make use of Prop. \ref{Prop4} to realize that \emph{the only basis (up to basis changes) in which the spontaneous decay is possible is the one which diagonalizes the free Hamiltonian}. This observation follows from noticing that as it is usually done in a phenomenological, though correct spirit to obtain decay we must add damping terms $-\gamma\rho_{\cdot\times}$ to the time evolution equation of each of the density matrix components $\rho_{\cdot\times}$ expressed in the free hamiltonian eigenvector basis (cf. \cite{Loudon}, page 64), which is equivalent to using a Lindblad operator of the form $$\matr{0}{0}{\gamma}{0}$$Following Prop. \ref{Prop4} this Lindblad operator can only be obtained when using the previous stochastic formalism if $H$ is diagonal, that is, in the free hamiltonian eigenvector basis. Another way of stating it is by noticing that it is the spontaneous emission which chooses the basis.

\bigskip

Finally the same hypotheses assumed in the orthodox quantum formalism (trace conservation, probability interpretation) are also taken up when the stochastic nature of quantum systems is considered.  

\bigskip

Note that the whole scheme is only based on a simple hypothesis: inherent randomness, the quantitative consequences of which are derived using just well-established physics-free stochastic methods.

\section{Conclusions}

In this paper we have considered the hypothesis of letting quantum systems show an inherent random nature. Mathematically we express this fact using a well-established theorem of decomposition of random variables using stochastic calculus. We then apply the usual definition of density operator for quantum systems being represented by stochastic state-vectors and impose normalization. As a result of these hypotheses we show how this stochastic evolution is determined by the usual hamiltonian $H$ of the system and a set of new operators $v_{k}(t)$ which characterizes the random behaviour of the evolution. The effect of these new elements in the description of the evolution of a quantum system makes possible to connect several originally independent facts. 

\bigskip

Firstly under suitable physical conditions, we may establish the general form of a completely positive Markovian evolution, i.e. a Lindblad evolution, the Lindblad operators being a more or less complicated expression of the hamiltonian $H$ and the operators $v_{k}(t)$. Secondly, choosing adequately  the operators $v_{k}(t)$ we may reproduce the phenomenon of spontaneous emission for a two-level quantum system in QED vacuum. Thirdly, choosing again these operators in a convenient way a system can show intrinsic decoherence, thus reducing the role of the observer in the measurement process. Finally, the formalism provides a natural way to attack the preferred basis problem, since now we have new phenomena such as spontaneous emission entering the physical picture which suggests a basis singularization.

\bigskip

Nevertheless some remaining points should also be addressed. First and most important the question of a physical principle which provides the form of the operators $v_{k}(t)$ must be investigated. In particular, in the same way as symmetry considerations usually provide a hint of the form of the hamiltonian $H$ of a system, it would be desirable to have some sort of similar reasoning to work out the form of $v_{k}(t)$ in each concrete situation. Second, though the principles of the scheme are well established above, a generalization to more than 2 energy-level systems is necessary. Work in both these directions is in progress.

\section*{Acknowledgments}
We acknowledge the support of Spanish Ministry of Science and Technology under project No. BFM2000-0013. One of us (D.S.) must also acknowledge the support of Madrid Education Council under grant BOCAM 20-08-1999.


\appendix
\section{Proof of Prop. \ref{ExpVal}}

The proof of proposition \ref{ExpVal} is a conjugation of Ito's formula and the polar decomposition of operators acting upon Hilbert spaces. Using (\ref{StEvolOper}) in (\ref{DensOp}) and making use of Ito's formula we get 

\begin{eqnarray}
\rho(t)&=&\mathbb{E}U_{st}(t)\rho(0)\mathbb{E}U_{st}^{\dagger}(t)+\nn\\
\label{DensOP}&+&\sum_{nm=1}^{N^{2}}a_{nm}\int_{0}^{t}v_{n}(s)\rho(0)v_{m}^{\dagger}(s)ds,
\end{eqnarray}
 where the correlation matrix $\mathbb{E}[W_{t}^{n}W_{t}^{m*}]=a_{nm}t$ has been taken into account.

\begin{table*}
\begin{tabular}{rcl}
$\Lambda_{11}=-|\ell_{21}|^{2}$ & &$\Lambda_{12}=ih_{12}+\frac{1}{2}[(2\ell_{11}-\ell_{22})\ell_{21}^{*}-\ell_{12}\ell_{11}^{*}]$ \\ 
$\Lambda_{13}=-ih_{21}+\frac{1}{2}[(2\ell_{11}^{*}-\ell_{22}^{*})\ell_{12}-\ell_{12}^{*}\ell_{11}]$& & $\Lambda_{14}=|\ell_{21}|^{2}$ \\ 
& & \\ 
$\Lambda_{21}=ih_{21}+\frac{1}{2}(\ell_{11}\ell_{12}^{*}-\ell_{21}\ell_{22}^{*})$ & & $\Lambda_{22}=-i(h_{11}-h_{22})+\frac{1}{2}[2\ell_{11}\ell_{22}^{*}-\sum_{i=1}^{4}|\ell_{ii}|^{2}]$ \\ 
$\Lambda_{23}=\ell_{21}\ell_{12}^{*}$ & & $\Lambda_{24}=-ih_{21}+\frac{1}{2}[-\ell_{11}\ell_{12}^{*}+\ell_{21}\ell_{22}^{*}]$\\ 
& & \\ 
$\Lambda_{31}=-ih_{12}+\frac{1}{2}[\ell_{11}^{*}\ell_{12}-\ell_{21}^{*}\ell_{22}]$ & & $\Lambda_{32}=\ell_{21}^{*}\ell_{12}$ \\ 
$\Lambda_{33}=i(h_{11}-h_{22})+\frac{1}{2}[2\ell_{11}^{*}\ell_{22}-\sum_{i=1}^{4}|\ell_{ii}|^{2}$ & & $\Lambda_{34}=ih_{12}+\frac{1}{2}[-\ell_{11}^{*}\ell_{12}+\ell_{21}^{*}\ell_{22}]$\\ 
& & \\ 
$\Lambda_{41}=|\ell_{12}|^{2}$ & & $\Lambda_{42}=-ih_{12}+\frac{1}{2}[\ell_{12}(2\ell_{22}^{*}-\ell_{11}^{*})-\ell_{21}^{*}\ell_{22}]$\\ 
$\Lambda_{43}=ih_{21}+\frac{1}{2}[\ell_{12}^{*}(2\ell_{22}-\ell_{11})-\ell_{21}\ell_{22}^{*}]$ & & $\Lambda_{44}=-|\ell_{12}|^{2}$
\end{tabular}
\caption{Matrix entries $\Lambda_{ij}$'s in $L[E_{n}]=\sum_{m=1}^{4}\Lambda_{nm}E_{m}$.\label{LamCoef}}
\end{table*}

\bigskip

Now forcing trace preservation $\tr\rho(t)=\tr\rho(0)$ and using the cyclic property of the trace we may write 

\begin{eqnarray}
\tr[\mathbb{E}U_{st}^{\dagger}(t)\mathbb{E}U_{st}(t)\rho(0)&+&\nn\\
\sum_{nm=1}^{N^{2}}a_{nm}\int_{0}^{t}v_{m}^{\dagger}(s)v_{n}(s)ds\rho(0)]&=&\tr\rho(0),\nn\\
& & 
\end{eqnarray}
  which if it is to be valid for all $\rho(0)$ implies 

\begin{equation}
\mathbb{E}U_{st}^{\dagger}(t)\mathbb{E}U_{st}(t)+\sum_{nm=1}^{N^{2}}a_{nm}\int_{0}^{t}v_{m}^{\dagger}(s)v_{n}(s)ds=I.
\end{equation}

From this and making use of the polar decomposition of $\mathbb{E}\Ust$, we conclude

\begin{equation}
\mathbb{E}\Ust=\mathcal{U}(t)\left(I-\sum_{nm=1}^{N^{2}}a_{nm}\int_{0}^{t}v_{m}^{\dagger}(s)v_{n}(s)ds\right)^{1/2},
\end{equation}

where $\mathcal{U}(t)$ is an arbitrary unitary operator which is expressed in the usual exponential form.

\section{Proof of Prop. \ref{Prop4}}
\label{AppendProp4}

The idea of the proof is to use the linearity of the space $\mathcal{M}_{2}(\mathbb{C})$. We will firstly focus upon the canonical basis $\{E_{11},E_{12},E_{21},E_{22}\}$ which we shall rename respectively as $\{E_{1},E_{2},E_{3},E_{4}\}$. Then we calculate $L[E_{k}]$ for each $k$. Denoting $$L[E_{n}]=\sum_{m=1}^{4}\Lambda_{nm}E_{m},$$ the results are shown in table \ref{LamCoef}.

\bigskip

We will agree that these 16 quantities configure the $4\times 4$ matrix $\Lambda$. By virtue of the linearity of $\mathcal{M}_{2}(\mathbb{C})$ , we write 

\begin{equation}
X=\sum_{n=1}^{4}x_{n}E_{n}\Rightarrow L[X]=\sum_{n=1}^{4}x_{n}L[E_{n}].
\end{equation}

 Then the condition $L[X]=0$ transforms into 

\begin{equation}\label{NullCond}
(x_{1}\ x_{2}\ x_{3}\ x_{4})\cdot\Lambda=(0\ 0\ 0\ 0).
\end{equation}
 
Now we must restrict the form of $X$, since it must be of the type $X=v\rho_{0} v^{\dagger}$, where $\rho_{0}$ is a density matrix (positive selfadjoint matrix with unity trace). Using again the linearity of  $\mathcal{M}_{2}(\mathbb{C})$, we may express $$v=\sum_{n=1}^{4}\beta_{n}E_{n}$$ Then $X$ may be written as $X=\sum_{n}\alpha_{n}E_{n}$ with

\begin{eqnarray}\nn
\alpha_{1}=\vefil{\beta_{1}}{\beta_{2}}\rho_{0}\vecol{\beta_{1}^{*}}{\beta_{2}^{*}} & &\alpha_{2}=\vefil{\beta_{1}}{\beta_{2}}\rho_{0}\vecol{\beta_{3}^{*}}{\beta_{4}^{*}}  \\ \nn
\alpha_{3}=\vefil{\beta_{3}}{\beta_{4}}\rho_{0}\vecol{\beta_{1}^{*}}{\beta_{2}^{*}} & &\alpha_{4}=\vefil{\beta_{3}}{\beta_{4}}\rho_{0}\vecol{\beta_{3}^{*}}{\beta_{4}^{*}}  \\ \nn
\end{eqnarray}

These $\alpha_{i}'s$ show the following immediate properties

\begin{enumerate}
\item $\alpha_{1}=0\quad\forall\rho_{0}\Leftrightarrow\beta_{1}=\beta_{2}=0\qquad\Rightarrow\alpha_{2}=0\quad\forall\rho_{0}$
\item $\alpha_{4}=0\quad\forall\rho_{0}\Leftrightarrow\beta_{3}=\beta_{4}=0\qquad\Rightarrow\alpha_{3}=0\quad\forall\rho_{0}$
\item $\alpha_{2}=0\quad\forall\rho_{0}\Leftrightarrow\alpha_{3}=0\quad\forall\rho_{0}$
\item $\alpha_{2}=0\quad\forall\rho_{0}\Rightarrow \beta_{1}=\beta_{2}=0 \textrm{ or }\beta_{3}=\beta_{4}=0$. In the first case we get $\alpha_{1}=0$ whereas in the second we get $\alpha_{4}=0$.
\end{enumerate}

Then, in matricial form, $X=v\rho_{0}v^{\dagger}$ may show the following aspects:

\begin{itemize}
\item If $v=\matr{\beta_{1}}{\beta_{2}}{0}{0}$, then $X=\alpha_{1}E_{1}$.
\item If $v=\matr{0}{0}{\beta_{3}}{\beta_{4}}$, then $X=\alpha_{4}E_{4}$.
\item If $v=\matr{\beta_{1}}{\beta_{2}}{\beta_{3}}{\beta_{4}}$ with at least three $\beta_{k}$'s not equal to $0$, then  $X=\sum_{n=1}^{4}\alpha_{n}E_{n}$.
\end{itemize}

Carrying these $\alpha_{i}$'s to (\ref{NullCond}) and taking into account the expression for $\Lambda_{ij}$'s we obtain the conditions announced in Prop. \ref{Prop4}.

\section{The Jaynes-Cummings model}

The Jaynes-Cummings hamiltonian can be straightforwardly diagonalized by noticing that in the $|\pm,n\rangle$ basis it adopts the matricial representation\\

\[\left(\begin{smallmatrix}
-\omega_{0}/2&0&0&\cdots&0&0&\cdots\\
0&\omega_{0}/2&\epsilon&\cdots&0&0&\cdots\\
0&\epsilon&-\omega_{0}/2+\omega&\cdots&0&0&\cdots\\
\vdots&\vdots&\vdots&\ddots&0&0&\cdots\\
 0 & \cdots & \cdots  & 0   & \omega_{0}/2+n\omega & \epsilon\sqrt{n+1} & 0 \\
0 & \cdots & \cdots  & 0  & \epsilon\sqrt{n+1}  & -\omega_{0}/2+(n+1)\omega & 0 \\
 0 & \cdots & \cdots & \cdots & 0 & 0 & \ddots
\end{smallmatrix}\right)\]

from which we extract the eigenvalues ($n\ge 0$)

\begin{eqnarray*}
E_{n+1}^{\pm}&=&(n+\frac{1}{2})\omega\pm\sqrt{\left(\frac{\omega-\omega_{0}}{2}\right)^{2}+(n+1)\epsilon^{2}}\\
E_{0}^{-}&=&-\omega_{0}/2
\end{eqnarray*}

 and the eigenvectors ($n\ge 0$)

\begin{eqnarray*}
|u_{n+1}^{+}\rangle&=&\phantom{-}\cos\theta_{n+1}|+,n\rangle+\sin\theta_{n+1}|-,n+1\rangle\\
|u_{n+1}^{-}\rangle&=&-\sin\theta_{n+1}|+,n\rangle+\cos\theta_{n+1}|-,n+1\rangle\\
|u_{0}^{-}\rangle&=&|-,0\rangle,
\end{eqnarray*}

where 

\begin{eqnarray*}
\tan\theta_{n+1}=\frac{\frac{\omega-\omega_{0}}{2}+\sqrt{\left(\frac{\omega-\omega_{0}}{2}\right)^{2}+(n+1)\epsilon^{2}}}{\epsilon\sqrt{n+1}}.
\end{eqnarray*}

These are the eigenvalues and eigenvectors of the dressed atom, which is commonly used in quantum optics (cf. \cite{Peng}). Notice that the set $\{|u_{n}^{\pm}\rangle,|u_{0}^{-}\rangle\}$ constitutes an orthonormal basis for the joint Hilbert space of atom+field.

\end{document}